\documentclass[runningheads]{llncs}
\usepackage[T1]{fontenc}
\usepackage{graphicx}
\begin{document}
\title{Exploring the Effects of Different Asymmetric Game Designs on User Experience in Collaborative Virtual Reality}
\titlerunning{Effects of Different Asymmetric Game Designs on UX in Collaborative VR}
%
\author{Francesco Vona\inst{1}\orcidID{0000-0003-4558-4989} \and
Evelyn Romanjuk\inst{1}\orcidID{0009-0003-8235-9120} \and
Sina Hinzmann\inst{1}\orcidID{0009-0000-2054-6623} \and
Julia Schorlemmer\inst{1}\orcidID{0009-0004-7388-9389} \and
Navid Ashrafi\inst{1}\orcidID{0009-0005-8398-415X} \and
Jan-Niklas Voigt-Antons\inst{1}\orcidID{0000-0002-2786-9262} }
\authorrunning{Vona et al.}
%
\institute{Hamm-Lippstadt University of Applied Sciences \\
\email{name.lastiname@hshl.de}}
\maketitle              
\begin{abstract}
The risk of isolation in virtual reality (VR) stems from the immersive nature of the technology. VR can transport users to entirely virtual environments, often disconnecting them from the physical world and real-life interactions. Asymmetric multiplayer options have been explored to address this issue and encourage social interaction by requiring players to communicate and collaborate to achieve common objectives. Nevertheless, research on implementing these designs and their effects is limited, mainly due to the novelty of multiplayer VR gaming. This article investigates how different game design approaches affect the player experience during an asymmetric multiplayer VR game. Four versions of a VR experience were created and tested in a study involving 74 participants. Each version differs in terms of the sharing of virtual environments (shared vs separated) and the players' dependency on the experience (mutual vs unidirectional).  The results showed that variations in game design influenced aspects of the player experience, such as system usability, pragmatic UX quality, immersion control, and intrinsic motivation. Notably, the player roles and the co-presence in the virtual environment did not simultaneously impact these aspects, suggesting that the degree to which players depend on each other changes the player experience.

\end{abstract}
\section{Introduction}
Virtual Reality (VR) has become a widely embraced technology, especially within video games, offering users an incredibly immersive experience. However, while VR games excite many players, they often receive criticism for potentially causing feelings of isolation and loneliness \cite{Boland2015}. As players wear the head-mounted display (HMD), they are transported into a virtual world yet simultaneously disconnected from their physical surroundings, leading to a solitary VR experience. A common strategy to address this issue is integrating other individuals into the gaming experience \cite{Karaosmanoglu2021}. This integration can vary from passive involvement, such as spectators watching the HMD player's perspective on a TV screen, to active participation in the game as a non-HMD player. Adding this social element enhances the multiplayer gaming experience, which is particularly favored among video game enthusiasts, who often spend considerable gaming time with online and offline friends, partners, or family members \cite{ESA2022}. Social play fosters community building and strengthening and promotes skills such as critical thinking, collaboration, creativity, and communication \cite{ESA2022,Gandolfi2018,Seif2010}.

Beyond addressing social isolation, enhancing the overall Gaming Quality of Experience (QoE) has become a growing area of interest and active standardization efforts \cite{Moeller2015}. Researchers explore various methods to improve the player’s engagement and immersion, including physiological techniques such as measuring brain activity to assess game-related states like flow \cite{Nunez2019}. Additionally, biofeedback mechanisms have been explored to dynamically adapt gameplay based on the player’s physiological responses, fostering a more interactive and immersive gaming experience \cite{Kojic2019}.

One intriguing way of designing multiplayer games involves employing an asymmetric approach, which capitalizes on disparities among players and devices, making it particularly relevant for addressing potential isolation in VR gaming. This study explores asymmetric game designs and their impact on User Experience (UX). Theoretical research findings in this area were applied to develop a functional multiplayer VR game - \textit{"Mission: Clear Candy"} The game was then used to assess the effects of asymmetric game designs on UX in an experiment involving 74 participants. The results of this study offer valuable insights into this specific gaming approach and how UX can be enhanced by incorporating a non-HMD player into a VR gaming experience.

\section{Theoretical Background}
This section presents a "best-fit" framework for asymmetric multiplayer VR games, outlining key components that underlie significant game design aspects. Additionally, it explores the definition of player experience (PX), both generally and within asymmetric VR games, providing insights into players' perceptions and the factors shaping their experiences.

\subsection{Best-Fit Framework}
The framework for asymmetric multiplayer VR games, developed by \cite{Rogers2021}, results from a comprehensive review of recent literature. It aims to explore different types of asymmetry and their effects on players. By analyzing the key ideas frequently referenced in the literature, the authors proposed the "best-fit" framework. This framework consists of four main components crucial in understanding important aspects of asymmetric multiplayer VR games.

The MDA framework by Hunicke et al. \cite{Hunicke} is the foundation of this proposed framework, integrating game design, development, analysis, and research into a structured methodology. It includes Mechanics, Dynamics, and Aesthetics (MDA), describing the elements of a game and their interactions. Mechanics involve fundamental elements manipulated by game designers, shaping the game experience, while Dynamics refer to responses to player inputs during gameplay. Aesthetics encompass various factors such as sensation, narrative, challenge, fellowship, discovery, expression, and submission. However, only the Aesthetics component was used in the creation of the best-fit framework.

Furthermore, the "best-fit" framework integrates a customized version of the MDA framework for designing asymmetric games, as suggested by \cite{Harris2016}. This variant focuses on improving asymmetric game designs by identifying ways to use asymmetry based on the core MDA framework. The Mechanics component identifies six types of asymmetry—ability, challenge, interface, information, investment, and goal/responsibility. Interdependence between players emerges as a key dynamic in asymmetric games, where players must rely on each other to achieve shared goals, resulting in various forms of interdependence—mirrored, unidirectional, and bidirectional. The framework also considers the manipulation of synchronicity and timing in gameplay, incorporating concepts like asynchronous, sequential, expectant, concurrent, and coincident timing. The Aesthetics of asymmetric games are explored using insights from a game study called "Beam Me ‘Round, Scotty!" \cite{Harris2015}. This exploration reveals aesthetics such as leadership, effects of player skill, familiarity with the interface, partner familiarity, interdependence, necessity, and coordination.

In addition to in-game asymmetry, the analysis by \cite{Rogers2021} emphasizes the importance of social asymmetry in gaming \cite{Kaye2016}. This involves shared physical spaces, age and ability differences, communication, teamwork, and knowledge disparities. The framework also includes research by \cite{Sykownik2018} on shared control, examining how this mechanic can affect players' experiences. Shared control involves game entities controlled by both players and requires communication to progress, aligning with interdependence. This framework was also the ground of this work and it was used as a reference to build the asymmetric prototype involved in the study. 

\subsection{Player Experience}
Like user experience, the term player experience is utilized in game user research to describe how a player perceives the interaction between themselves and the game \cite{Gerling2011}. Several researchers have attempted to develop models to describe the components representing PX. However, there is currently no universally agreed-upon definition of PX, and there is limited understanding of which components are the most important \cite{Nacke}.

A frequently cited model in the context of PX is GameFlow by Sweetser and Wyeth \cite{Sweetser2005}, used to evaluate player enjoyment in gaming. It is based on Csikszentmihalyi's flow concept \cite{Csikszentmihalyi2009}, which identifies universal factors of an enjoyable experience and is applied in various fields. Other models, such as the Game Experience Questionnaire (GEQ), also capture essential facets of PX, including competence, sensory immersion, flow, tension, challenge, negative and positive affect, and social aspects \cite{IJsselsteijn2013}, though it has been criticized for its unstable factor structure and reliability concerns in certain subscales \cite{Law2018}. Wiemeyer et al.'s psychological model combines generic and domain-specific models, with the latter specifically applying to gaming \cite{Wiemeyer2016}. These models encompass theories such as self-determination, presence, immersion, and GameFlow, providing a comprehensive framework.

PX considerations extend to specific fields like asymmetric multiplayer and VR games. Studies have examined PX aspects through various methods. Karaosmanoglu et al. designed an asymmetric VR game to explore different interdependence types among players, assessing affective state, immersion, presence, and enjoyment \cite{Karaosmanoglu2021}. Gugenheimer et al. evaluated their co-located VR game "ShareVR" using GEQ and a presence questionnaire \cite{Gugenheimer2017}. Emmerich and Masuch conducted a study on the impact of game design patterns on player interaction, using questionnaires to analyze PX and social engagement \cite{Emmerich2017}.

Certain aspects hold repeated significance within multiple PX models. Immersion is a crucial component discussed across various models, encompassing technical attributes and psychological engagement \cite{Sweetser2005,Slater1996}. Social interaction, especially in multiplayer contexts, is a recurrent focus, with studies highlighting heightened engagement when playing against another person instead of a computer opponent \cite{Ravaja2006}. Based on these studies, frameworks for understanding game settings' social aspects have also emerged \cite{DeKort2008}.

Recurring aspects of player experience, such as immersion and social interaction, play a crucial role in assessing PX across different game settings. These factors help elucidate how various gameplay elements influence player engagement. By integrating them into UX assessment—which offers a broader perspective on user satisfaction and overall interaction—a more holistic understanding of the player’s complete experience can be achieved. This approach combines both game-specific elements with general user experience factors, providing a comprehensive framework for evaluating engagement in diverse gaming contexts.

\section{Methods}
\subsection{Mission: Clear Candy}
To investigate how different asymmetric game designs affect UX, an asymmetric multiplayer game called "\textit{Mission: Clear Candy}" was developed, using the "best-fit" framework as a foundation for game design choices.
In \textit{Mission: Clear Candy}, two players have to explore the laboratory of a candy factory and take a sample of a mysterious hidden substance. One of the players takes the role of a secret agent and plays the game in VR, using an HMD and two VR controllers, while the other person plays as a hacker, using a desktop PC, a keyboard, and a mouse. While the HMD player explores the 3D level and can interact with machines and other laboratory objects, the non-HMD player can navigate through a fictional desktop containing folders and files with the information needed to complete the mission or alternatively control a robot in the same 3D world. Thus, the players have their own game worlds, abilities, and pools of information and must collaborate to solve the puzzles and overcome the presented obstacles. As they lack insight into each other's game world, constant verbal communication is necessary to achieve their goal.

\subsection{Study Design}
An experiment was run to investigate how UX changes with different implementations of the asymmetric VR game. A between-subjects design was chosen with four variations of \textit{Mission: Clear Candy}. Since the game's puzzles were mostly the same for each version, participants needed to be unfamiliar with the game. Therefore, they were randomly assigned to only one of the versions. The same questionnaires were administered to all participants.

\subsection{Apparatus}
Each version of \textit{Mission: Clear Candy} varies in terms of player roles, interactions, and dependencies. In the following, a summary of the different versions is provided:
\begin{itemize}
    \item \textbf{Version A (Separated Environments, Bidirectional Player Role Dependency)}: In version A of the game, the HMD player navigates as an agent through the laboratory in VR, while the non-HMD player assumes the role of a hacker in their own game world, presented as a 2D interface. This interface includes additional tools and information necessary to progress through the game, which are inaccessible to the agent. As a result, there is an asymmetry in abilities and information between the player roles. This necessitates constant verbal communication between the players, as they cannot directly see or interact with each other within the game. Additionally, a bidirectional type of dependence is forced in this version. This means that in some instances, the agent will rely on the hacker's information or actions, while in others, the hacker will depend on the agent. For example, when players need access to the candy machine presented in the game, the hacker can provide the password for an employee card but requires information from the agent to determine the correct card.
    \item \textbf{Version B (Separated Environments, Unidirectional Player Role Dependency)}: In version B, players operate within separated environments similar to version A. However, this version establishes a unidirectional dependency, making the agent more reliant on the hacker than vice versa. This places the non-HMD player in a leading role within the game design, as they possess more information to solve puzzles and must coordinate both their actions and the agent's. For example, when players need to find a recipe code, only the hacker possesses the necessary information and interface to input the code, but they lack one crucial piece of the puzzle, which the agent in the laboratory can provide. Consequently, the non-HMD player must utilize all available information and tools to determine the next steps and effectively communicate their thoughts to the HMD player to obtain missing information.
    \item \textbf{Version C (Shared Environment, Unidirectional Player Role Dependency)}: In this version, both players share the same game world, aiming to assess how UX aspects change when players can see and interact with each other compared to previous versions. To achieve this, the non-HMD player controls a robot within the laboratory, capable of scanning various objects for information. The HMD player can also carry the robot through the environment, similar to non-playable objects. In version C, the dynamic of version B is reversed, making the hacker more reliant on the agent who assumes the leading role. For example, while searching for a valid employee ID card and password, the agent requires the robot to scan a card for the respective password, allowing them to enter the password into the candy machine independently. However, the robot often needs the agent's assistance to reach cards, as it cannot access elevated objects independently.
    \item \textbf{Version D (Shared Environment, Bidirectional Player Role Dependency)}: Finally, version D establishes a bidirectional dependency, where players rely on each other simultaneously, necessitating close cooperation and clear communication without a predefined hierarchy. For instance, some puzzles are redesigned to require simultaneous input from both players to achieve a goal, such as opening a door by pressing two buttons simultaneously. Similar to version C, the non-HMD player operates within the same 3D level as the HMD player, facilitating interaction between them.

\end{itemize}
To develop a functional application of \textit{Mission: Clear Candy}, the game was created using the Unity 2021.3.15f1 game engine. The VR components of the game were implemented using the XR Interaction Toolkit 2.0.4 and the OpenXR Plugin 1.5.3. The Meta Quest 2 HMD and its corresponding controllers were utilized for development and subsequent testing. For enhanced performance, the HMD was connected to a PC via cable, on which the game was executed.

\subsection{Dependent Variables}
As outlined in the previous section, there is no strict, standardized method for measuring PX, which is why we did not include PX-specific questionnaires in our study. Instead, we selected UX questionnaires based on existing literature, as they provide well-established and validated measures for assessing various UX components. To ensure a more comprehensive evaluation, we expanded beyond traditional UX measures by incorporating additional questionnaires related to intrinsic motivation, immersive experience, and social presence. These measures, as highlighted in the previous section, represent a common intersection between UX and PX, allowing us to capture a broader range of experiential factors while maintaining methodological reliability. 
\paragraph{PRE-GAME QUESTIONNAIRES} At the beginning of the experiment, the participants were asked to fill in a pre-game questionnaire, which consisted of two parts.
\begin{itemize}
    \item Demographic questions, asking for gender, age, profession, the average time spent gaming per week in hours, and the hours spent using a VR headset with controllers.
    \item Affinity for Technology Interaction (ATI) scale: Used to gain insight into the participants’ attitudes towards technical systems and their interaction with such systems. This scale consists of 9 items on a 6-point Likert scale (1 = completely disagree, 6 = completely agree)\cite{Franke2019}. 
\end{itemize}
\paragraph{POST-GAME QUESTIONNAIRES} After the experiment, the participants were asked to fill in a post-game questionnaire with multiple questionnaires that addressed various UX and PX aspects.
\begin{itemize}
    \item System Usability Scale (SUS): Measured overall usability perception. It consists of 10 items on a 5-point Likert scale (1 = strongly disagree, 5 = strongly agree) \cite{Brooke1995}. The dependent variable SUS was derived from this questionnaire. 
    \item User Experience Questionnaire - Short Version (UEQS): Assessed pragmatic and hedonic user experience qualities. The questionnaire contains eight items on a 7-point Likert scale, each presenting two opposing adjectives (1 = obstructive, complicated, inefficient, confusing, boring, not interesting, conventional, usual, 7 = supportive, easy, efficient, clear, exciting, interesting, inventive, leading edge)\cite{Schrepp2017}. This leads to three dependent variables UEQS Pragmatic, UEQS Hedonic, and UEQS Overall, which includes all eight items.  
    \item Intrinsic Motivation Inventory (IMI): Evaluated subjective experience and intrinsic motivation. It comprises 22 items on a 7-point Likert scale (1 = not at all true, 7 = very true). The items are evaluated on multiple subscales \cite{Ryan2000}. Four dependent variables, IMI Interest, IMI Competence, IMI Choice, and IMI Pressure, could be obtained from these subscales.
    \item Immersive Experience Questionnaire (IEQ): Measured immersive qualities in various aspects. Thirty-one items on a 7-point Likert scale (1 = not at all, 7 = a lot), as well as one single question to evaluate the immersion overall (\mbox{1 = not} at all immersed, 10 = very immersed) \cite{Jennett2008}.  The dependent variables IEQ Challenge, IEQ Control, IEQ Dissociation, IEQ Emotional, IEQ Cognitive, and IEQ Overall were obtained from this.    
    \item Competitive and Cooperative Presence in Gaming Questionnaire (CCPIG): Measured social presence between players \cite{Hudson2014}. The section presents 25 items on a 5-point Likert scale (1 = completely disagree, 5 = completely agree) and is grouped in the subscales of team identification, social action, motivation, and team value. These subscales lead to the dependent variables CCPIG Identification, CCPIG Action, CCPIG Motivation, and CCPIG Value.
\end{itemize}

\subsection{Independent Variables}
Three main independent variables were considered:
\begin{itemize}
    \item \textbf{Role}: Each game version features two roles – Hacker and Agent – different in their abilities and responsibilities. They also use two different input devices, desktop PC (Hacker) and HMD (Agent), varying in their immersion degree. Roles are chosen by the players at the beginning of the experience. These variables were chosen to explore how using devices with varying degrees of immersion impacts UX in an asymmetric game.
    \item \textbf{Environment}: Explored whether players played in shared or separated virtual environments (Shared, Separated). In the Game Versions C and D, the non-HMD player navigates through the same 3D level as the HMD player, making it possible for them to interact with each other more closely (Shared), whereas in versions A and B, the non-HMD player has their game world, separating them from their teammate (Separated). These variables were chosen to explore how sharing a virtual environment impacts UX in an asymmetric game.
    \item \textbf{Game Version}: Refers to the different game versions (A, B, C, D). The game versions can be identified as a manipulation factor, as each game version presents a different and specific game design approach, combining the independent variables listed before. Additionally, the type of player dependency (unidirectional, bidirectional) varies between versions.
\end{itemize}
\subsection{Participants}
In total, 74 people participated in the experiment, resulting in 37 teams of two players. Forty-four participants (59.5\%) identified as male, 28 identified as female (37.8\%), and 2 (2.7\%) preferred not to disclose their gender. The average age was 25.23 years. When asked about their profession, 52 participants (70.3\%) stated they were students, 17 (23\%) would be researchers at a university, and 5 (6.8\%) would be employed elsewhere. On average, the participants reported spending 11.39 hours per week on gaming, with a median time of 6.5 hours (\textit{SD} = 15.44). In addition, the participants were asked about their experience with VR technology. Most of them had spent little to no time using an HMD. Concerning affinity to technology, the analysis on the ATI scale yielded a Cronbach’s alpha of .83, which shows the high internal consistency of the answers the participants gave. The average ATI score was 4.37 (\textit{SD} = 0.79), which indicates a moderate level of affinity for technology among the participating players.
\subsection{Experiment Procedure}
\begin{figure}[ht]
  \centering
  \includegraphics[width=\linewidth]{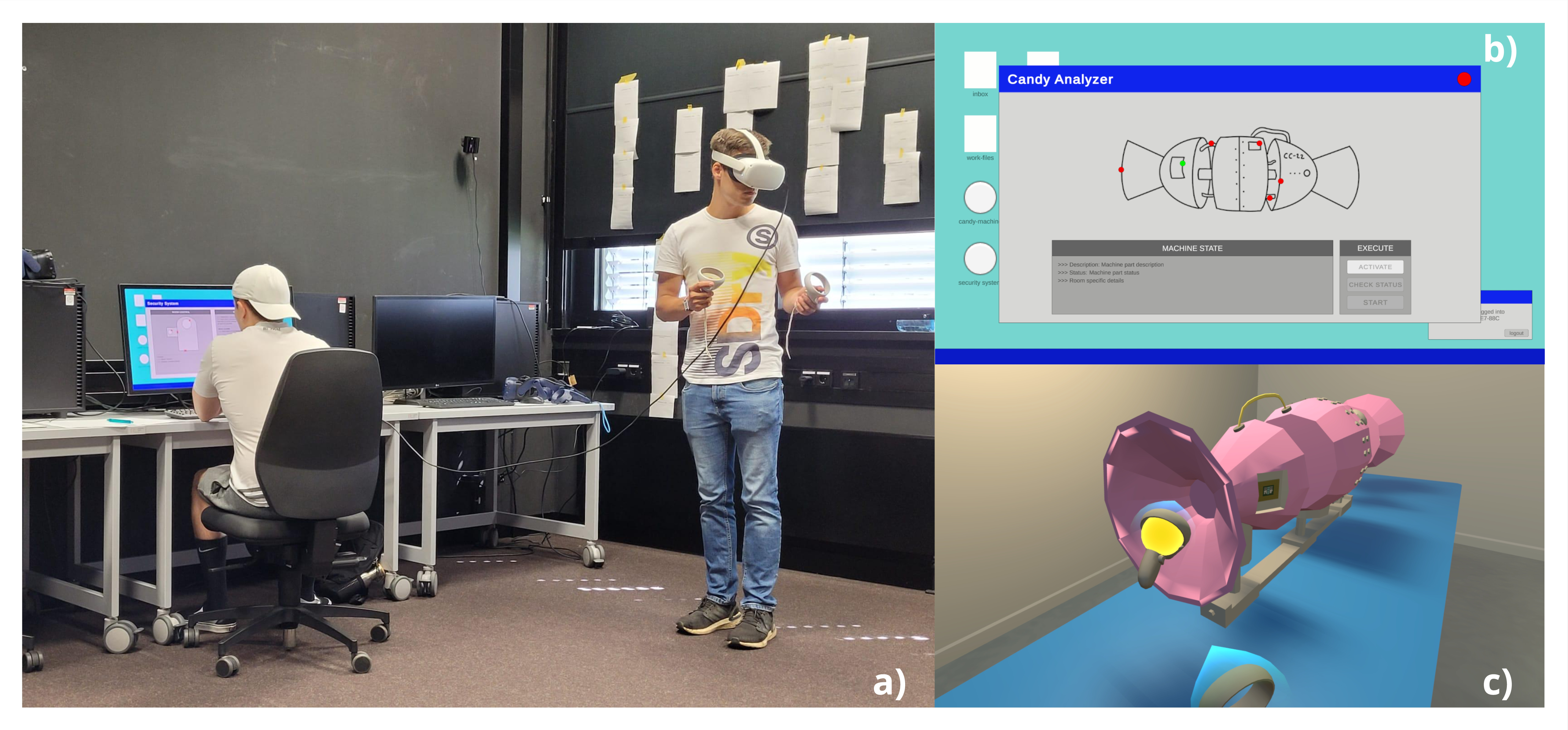}
  \caption{a) Study settings: Two participants playing \textit{Mission: Clear Clandy}. b) A view of the game from the non-HMD point of view. c) A view of the game from the HMD point of view. }
  \label{fig:MCC-study}
\end{figure}

Experiments were conducted in a controlled environment, specifically in one room at the university campus. The procedure lasted approximately 90 minutes, and participants received 25 euros as compensation. This study received approval from the University Ethics Commission. Throughout the experiment, a desktop computer was utilized for running the game and for the non-HMD player to operate (Fig. \ref{fig:MCC-study}). The Meta Quest 2 headset was connected via cable for better performance. Additionally, a second desktop computer was provided to complete the questionnaires.

Participants were introduced to the study, provided consent forms, assigned a version, and asked to complete pre-game questionnaires. They were given the freedom to choose their roles themselves. Following instructions, participants engaged in gameplay for up to an hour. They were informed that they could stop the experiment at any point and were encouraged to communicate any discomfort or questions regarding the game controls to the experimenter during the session. If the team encountered difficulties progressing in the game, the experimenter offered hints or advice.

Subsequently, post-game questionnaires were completed. At the end of the experiment, 20 participants played version A of the game, while versions B, C and D were played by 18 participants each.

\section{Results and Discussion}
\begin{figure}[ht]
  \centering
  \includegraphics[width=\linewidth]{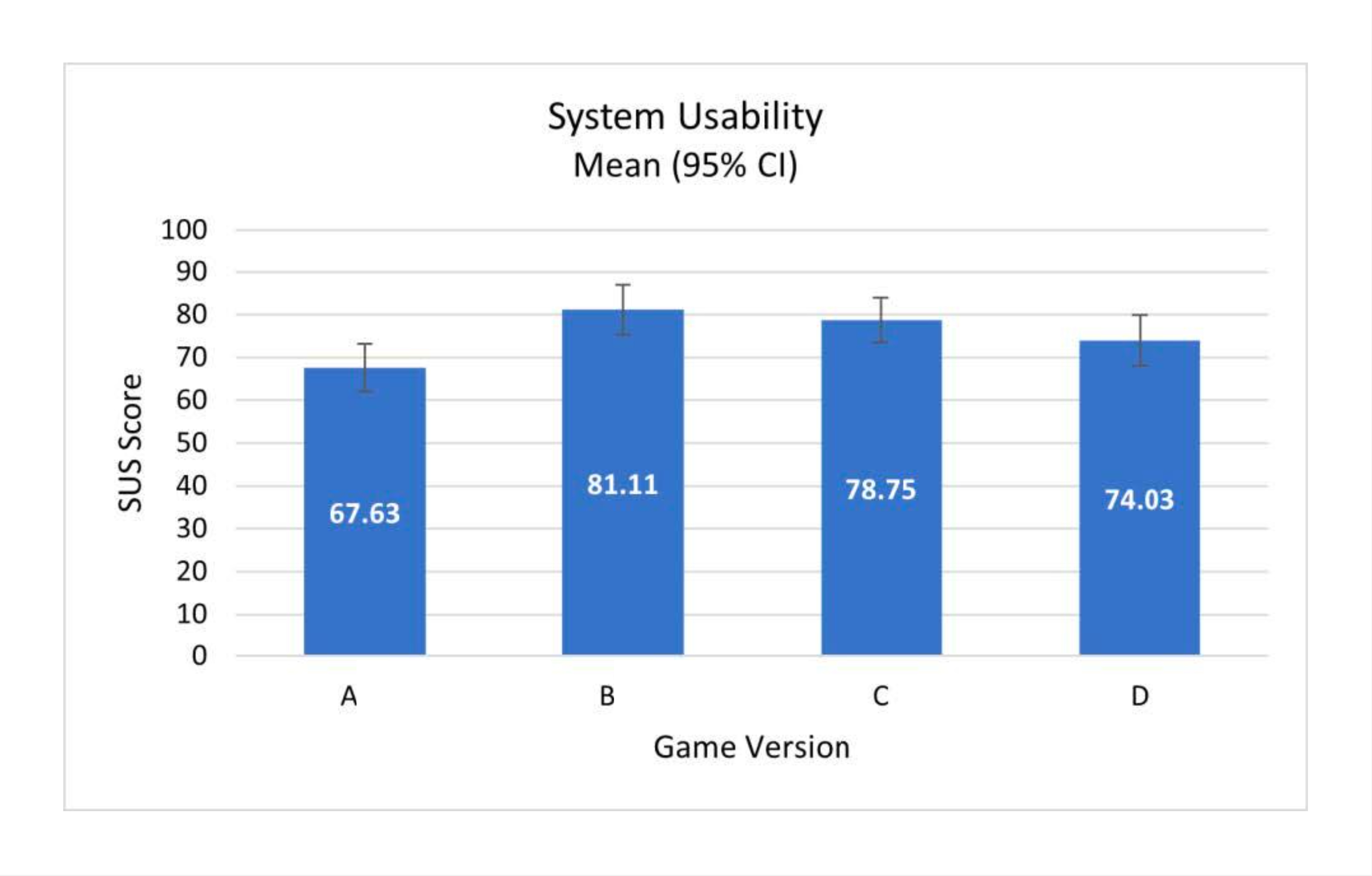}
  \caption{Comparison of average values of the SUS questionnaire for each Version. Versions: A = separated environments, bidirectional player role dependency; B = separated environments, unidirectional player role dependency; C = shared environment, unidirectional player role dependency; D = shared environment, bidirectional player role dependency.}
  \label{fig:sus-results}
\end{figure}

\paragraph{System Usability.}The participants' collective response yielded an average SUS score of 75.17 (\textit{SD} = 13.19). With the one-way analysis of variance (ANOVA) and the independent samples \textit{t}-tests, the effects of Game Versions, Roles, and Environments on system usability were examined. The ANOVA analyses revealed significant differences only between the Game Versions (\textit{F(}3, 70\textit{)} = 4.43, \textit{p} = .007, $\eta^2$ = .16), while no significant results were found with t-tests for the independent variables Role and Environment (Fig. \ref{fig:sus-results}). The Tukey HSD post-hoc test was run to compare the mean differences between each version. In the comparisons, Version A scored lower than Version B (\textit{p} = .007, mean difference = 13.49) and lower than Version C (\mbox{\textit{p} = .04}, mean difference = 11.1). There were no significant differences in system usability for other version combinations (\textit{p }> .05).

\begin{figure}[ht]
  \centering
  \includegraphics[width=\linewidth]{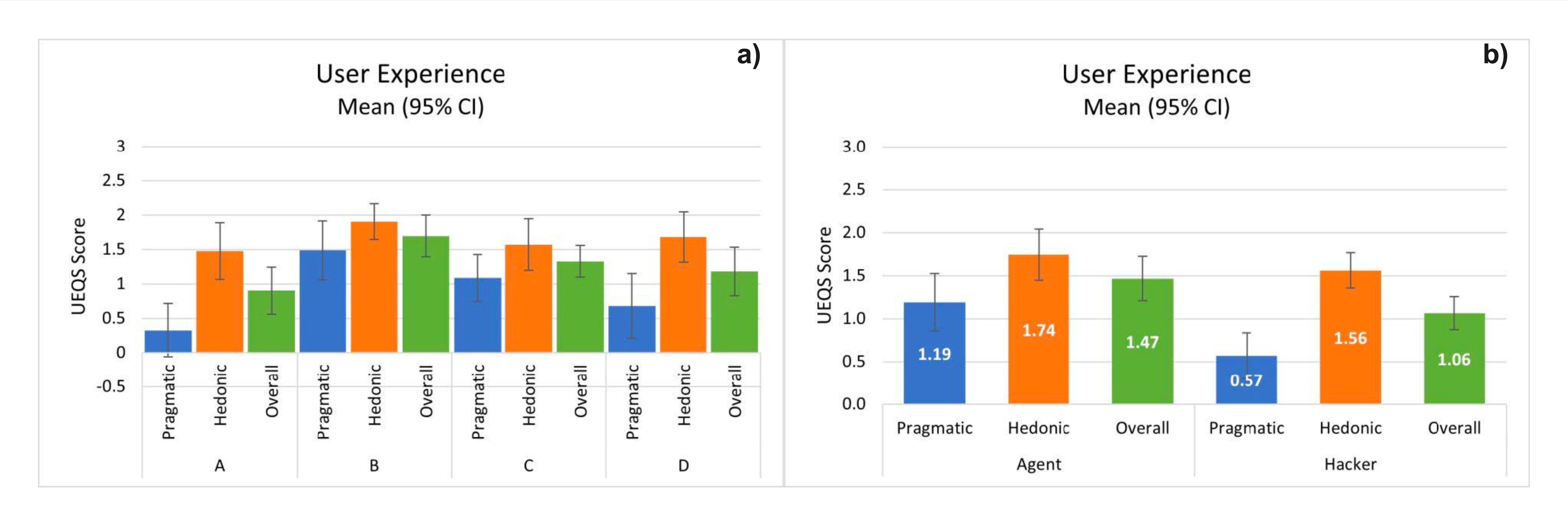}
    \caption{Comparison of average values of the UEQS questionnaire for each Version (a) and each Role (b). Versions: A = separated environments, bidirectional player role dependency; B = separated environments, unidirectional player role dependency; C = shared environment, unidirectional player role dependency; D = shared environment, bidirectional player role dependency.}
  \label{fig:ueq-results}
\end{figure}

\begin{figure}[ht]
  \centering
  \includegraphics[width=\linewidth]{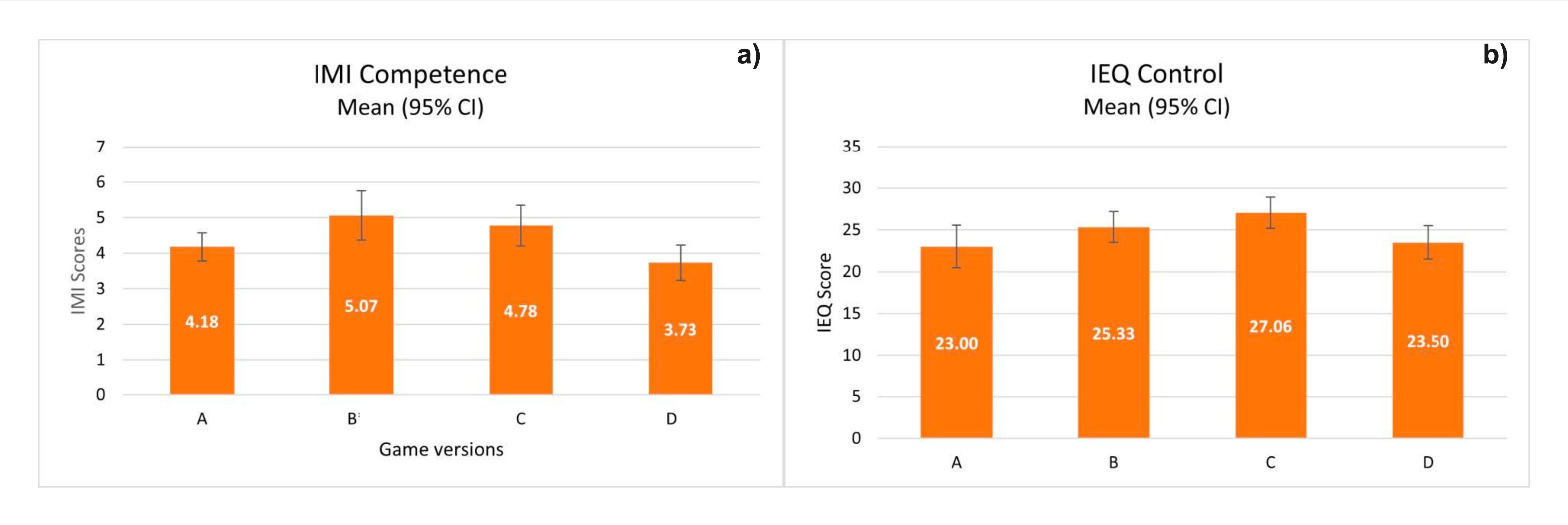}
    \caption{Comparison of average values of the IMI Competence (a) and IEQ Control (b) components for each Version. Versions: A = separated environments, bidirectional player role dependency; B = separated environments, unidirectional player role dependency; C = shared environment, unidirectional player role dependency; D = shared environment, bidirectional player role dependency.}
  \label{fig:imi+ieq-results}
\end{figure}

\paragraph{User Experience.} Over all participants, an average UEQS score of \textit{M} = 1.27 (\textit{SD} = 0.73) was calculated. The value of the pragmatic quality is \textit{M} = 0.88 (\textit{SD} = 0.98, $\alpha$ = .68) while a value of \textit{M} = 1.65 (\textit{SD} = 0.79,  $\alpha$ = .71) was calculated for the hedonic quality (Fig. \ref{fig:ueq-results}). The ANOVA results indicated that pragmatic quality (UEQS Pragmatic; \mbox{\textit{F}(3, 70) = 5.90}, \mbox{\textit{p} = .001}, \mbox{$\eta^2$ = .20}) and overall quality (UEQS Overall; \mbox{\textit{F}(3, 70) = 4.39}, \textit{p} = .007, $\eta^2$ = .16) exhibited significant variations across the Game Versions (Fig. \ref{fig:ueq-results} - left). The Tukey HSD post-hoc analysis gave more insight into which pairs of Game Versions differed significantly. Regarding UEQS Pragmatic, such differences were identified between versions A and B (\textit{p} < .001) as well as between versions B and D (\mbox{\textit{p} = .04}), while the analysis for UEQS Overall yielded a significant difference between versions A and B (\textit{p} = .004). The player Roles were compared using the independent samples \textit{t}-test. As found in the previous analysis for the Game Versions, the \textit{t}-test indicates significant differences between the UEQS Pragmatic scores of the Roles (\textit{t}(72) = 2.85, \textit{p} = .006) and between the UEQS Overall scores (\textit{t}(72) = 2.44, \textit{p} = .017) (Fig. \ref{fig:ueq-results} - right). The analysis with the independent variable Environment showed no significant difference between the conditions.
\paragraph{Intrinsic Motivation.} The mean IMI scores of overall participants were calculated for each subscale. The average score for IMI Interest is \textit{M} = 5.93 (\textit{SD} = 0.76), for IMI Competence, it is \textit{M} = 4.43 (\textit{SD} = 1.28), IMI Choice resulted in \textit{M} = 5.44 (\textit{SD} = 1.02) while the average score for IMI Pressure is at \textit{M} = 3.05 (\textit{SD} = 1.14). 
The one-way ANOVA test between the versions yielded a value of  \textit{F}(3, 70) = 4.57, \textit{p} = .006 and \(\eta^2\) = .16 for the subscale IMI Competence (Fig. \ref{fig:imi+ieq-results} - left).  The Tukey HSD posthoc test indicated a significant difference between versions B and D (\textit{p} = .007) as well as between versions C and D (\mbox{\textit{p} = .05}, which is just at the threshold). The analysis with the independent variable Roles and Environment showed no significant difference between the conditions.
\paragraph{Immersive Experience.} The calculated average IEQ Overall score was \textit{M} = 158.86 (\textit{SD} = 19.69).
The one-way ANOVA test, conducted to compare the immersion subscales between the Game Versions, showed only a significant difference for the values of IEQ Control ({\textit{F}(3, 70) = 2.95}, \mbox{\textit{p} = .038}, \mbox{$\eta^2$ = .11}) (Fig. \ref{fig:imi+ieq-results} - right). While the independent samples \textit{t}-test showed a significant difference in IEQ Dissociation scores between the Player Roles (\textit{t}(72) = 2.40,  \textit{p} = .02), no significant differences could be found between the two different Environments for any subscales.
\paragraph{Cooperative Presence.} The mean CCPIG scores overall participants were calculated for each subscale. The average score for CCPIG Identification is \textit{M} = 4.39 (\textit{SD} = 0.59); for CCPIG Actions, \textit{M} = 4.27 (\textit{SD} = 0.51); for CCPIG Motivation with \textit{M} = 4.04 (\textit{SD} = 0.57), and CCPIG Value with \textit{M} = 4.43 (\textit{SD} = 0.47). Game Versions, Player Roles, and Environments do not significantly affect cooperative social presence.
\section{Discussion}
\paragraph{System Usability}
The overall SUS score (M=75.17) indicates "good" usability\cite{Bangor2009} showed by a generally satisfactory level of usability across the different conditions. However, Game Versions showed significant differences in usability perception. Version A had lower usability ratings than version B and C, possibly due to alternating unidirectional player dependencies, leading to more communication challenges. 
Specific puzzle designs might also have influenced usability, as some puzzle variations required more advice from the experimenter, especially in versions with closer cooperation. Not getting enough feedback from the system or clear instructions on how to use specific controls might have led to confusion and, thus, lower scores regarding perceived system usability.
\paragraph{User Experience}
The mean score over all participants was calculated for the three dependent variables UEQS Pragmatic (\textit{M}=0.88), UEQS Hedonic (\textit{M}=1.65), and UEQS Overall (\textit{M}=1.27). Compared to the benchmark data, the pragmatic quality of the game in terms of UX is “below average”. The hedonic quality, on the other hand, can be interpreted as “excellent”. This results in a value for the UEQS Overall which can be interpreted as an “above average” quality of UX for the game across all conditions and participants \cite{Schrepp2017}. 
The one-way ANOVA test was applied to compare the Game Versions regarding the dependent variables. There is a significant difference between the Game Versions regarding the pragmatic qualities of UX. On the other hand, the results do not point towards significant differences for the hedonic qualities. Therefore, it can be concluded that the significant differences between the versions for UEQS Overall mainly stem from differences in the pragmatic aspects, while the hedonic UX elements are more consistent across the Game Versions. The Tukey HSD post-hoc analysis gave more insight into which pairs of Game Versions differed significantly and identified such differences between versions A and B as well as between versions B and D. These findings confirm the results on usability, where version A shows significantly lower system usability scores compared to version B. As the pragmatic qualities of UX also cover usability aspects, including attributes such as “complicated” or “efficient,” this would reinforce the conclusions made for usability, namely that player dependencies and specific puzzle implementations are possible reasons for the varying usability perception across the Game Versions. 
Further, the player roles were compared using the independent samples \textit{t}-test. As found in the previous analysis for the Game Versions, the \textit{t}-test indicates significant differences between the UEQS Pragmatic scores of the roles and the UEQS Overall scores. As before, it can be assumed that the differences in pragmatic quality lead to the differences overall. Thus, a player's role in the game may influence how the pragmatic UX quality is perceived, with playing as the Hacker presumably leading to a lesser perceived experience than playing as the Agent. On the other hand, the analysis with the independent variable Environment showed no significant difference between the conditions, so it might not make a difference in terms of UX whether the players play in a shared environment or separately. 
In summary, pragmatic UX differences were associated with Game Versions and Player roles. Version A's lower system usability scores align with the pragmatic UX findings, emphasizing the importance of usability in cooperative games. Designing games with clear, user-friendly tools for player actions is crucial, particularly in cooperative scenarios.
\paragraph{Intrinsic Motivation}
The analysis across all conditions resulted in moderately high mean values for IMI Interest (\textit{M}=5.93) and IMI Choice (\textit{M}=5.44). The results of the subscales IMI Choice and IMI Pressure support the assumption of an overall moderately high level of intrinsic motivation. For the subscale IMI Competence, the one-way ANOVA test has shown significant differences between the Game Versions. The \textit{t}-tests for the independent variable role showed no significant differences. Furthermore, no significant differences were found in the comparison between the two environments for any of the aspects of intrinsic motivation. Similarly to the findings for system usability, this leads to the assumption that the differences in perceived competence between versions could stem from factors other than Roles and Environments, such as the type of player dependency and the resulting way of communication between the players.

\paragraph{Immersive Experience}
The calculated average IEQ score (\textit{M}=158.86) shows a moderately high level of immersion. The one-way ANOVA test, conducted to compare the immersion subscales between the Game Versions, showed only a significant difference in the values of IEQ Control. While the independent samples \textit{t}-test showed a significant difference in IEQ Dissociation scores between the player roles, no significant differences could be found between the two different environments for any of the subscales. As the difference between IEQ Control values cannot be found with either of the \textit{t}-tests, it appears that neither the role nor the environment affected the perception of control. As in the previous analyses, this would lead to the assumption that the observable differences are due to the different kinds of player dependency and the resulting communication style. As control can be defined as “a sense of control over [one’s] actions” \cite{Sweetser2005}, its varying perceptions by the players across Game Versions could also be related to the possible usability issues discussed in the previous sections, as it could be argued that if a system does not provide sufficient usability, it may hinder the sense of control as well. However, this cannot be concluded from the data alone and would require further testing to see how these version-specific differences in UX aspects are related. 
In addition to the differences in perceived control, the second \textit{t}-test found a significant difference between the roles in terms of real-world dissociation. This subscale is concerned with the participants’ awareness of time and their physical surroundings and was found to result in higher scores for the agent role compared to the hacker. Here, it is likely that the greater dissociation originates from wearing an HMD in the Agent condition, as participants in this role cannot see their real-world environment while playing the game and may be less likely to be distracted by it. In contrast, the non-HMD player in the role of the hacker is playing on a desktop PC and can see their surroundings, as well as the playing HMD player due to the co-located setup. 

\paragraph{Cooperative Presence}
Across all participants and conditions, each subscale shows moderately high to high scores, with an average of M > 4. This indicates that the players generally felt a rather strong social presence during cooperative play.
Looking at the individual conditions, no significant differences between Game Versions, Roles, or Environments could be found, which leads to the conclusion that all of the implemented asymmetric game design variants can generate a relatively high social presence among the players. This could also be observed during experiments, as many of the participants talked a lot and shared their observations. In versions C and D, some of the players also visibly enjoyed interacting with each other, as they shared the same virtual environment.
Overall, every condition showed high levels of social presence, which were not influenced by the applied game designs. It could be assumed that the asymmetric multiplayer setting leads to an enjoyable experience in terms of social interaction, for example, because both players have a specific role and are required to rely on each other. 

\section{Conclusion}
This study investigated the effects of various game design options in asymmetric multiplayer VR games by conducting an experiment in which four different versions of a game were played and evaluated regarding PX and UX aspects. The Game Versions (A = separated environments, bidirectional player role dependency; B = separated environments, unidirectional player role dependency; C = shared environment, unidirectional player role dependency; D = shared environment, bidirectional player role dependency) featured two different Roles (Hacker, Agent), Environments (Separated, Shared), and types of player dependencies (unidirectional, bidirectional), which were compared. The study results have shown that the game design variations led to changes in multiple aspects in four out of five PX and UX categories, namely \textit{system usability}, \textit{pragmatic UX quality}, \textit{control} concerning immersion, and \textit{competence} as part of intrinsic motivation. It could also be found that in these cases, both variables Role and Environment were not affecting these aspects simultaneously, which led to the hypothesis that the type of player dependency, which changed between Game Versions, was a possible factor for changes in PX and UX. While the player roles were affecting the pragmatic quality of UX and the degree of real-world dissociation, the environments were not found to significantly affect PX and UX at all. Lastly, none of the independent variables had any significant effect on cooperative social presence or the enjoyment aspect of intrinsic motivation, which simultaneously resulted in moderately high to high scores on average.
These findings indicate that asymmetric multiplayer VR games have the potential to be generally enjoyable for both non-HMD and HMD players and to create a high social presence. In addition, the choice between a shared or separated environment for the players to operate in may not significantly impact PX, so the decision on which one to use in a game may depend more on the specific game mechanics than on PX concerns. Finally, special attention may be necessary when designing games featuring high player dependency regarding usability. To provide an enjoyable experience for both players, it is crucial to design game mechanics and interfaces that are not too confusing or complicated to use. This way, players may be able to focus more on the game experience itself and have an easier time working together toward a common goal.

\subsection{Limitations and Future Work}
The experiment was conducted with 74 participants, which is a relatively small sample size for a between-subject study. This is even more true for the subsample sizes, as some groups included only 18 participants. Thus, the results presented have to be treated with caution. Furthermore, the observations during the experiments and the player feedback showed possible flaws in specific puzzle implementations, especially in the versions that required closer cooperation between the players. Such puzzles were often perceived as confusing, as not enough feedback or instructions were given to the players in-game, which may have distorted the players' UX evaluations. However, these results can still be considered reliable given the consistency obtained in the data from the different questionnaires (SUS, UEQS). Based on the results and limitations of this study, it may be interesting to work on an iteration of \textit{Mission: Clear Candy}, in which puzzle implementations are improved to investigate if the effects on usability persist or if new insights can be made when players can delve more into the game without relying on external help. Additionally, focusing on player dependency may reveal more information about the effects of possible dynamics between a non-HMD and an HMD player, leading to more design possibilities for future games.

\begin{credits}
\subsubsection{\ackname} This work was supported by the European Union's Horizon Europe program under grant number 101092875 ``DIDYMOS-XR'' (https://www.didymos-xr.eu).

In this paper, we used Overleaf’s built-in spell checker, the current version of ChatGPT (GPT 4.0), and Grammarly. These tools helped us fix spelling mistakes and get suggestions to improve our writing. If not noted otherwise in a specific section, these tools were not used in other forms.
\end{credits}

%
%
%
%

\end{document}